\documentclass{jfm}

\usepackage{natbib}
\usepackage{hyperref}
\hypersetup{
    colorlinks = true,
    urlcolor   = blue,
    citecolor  = black,
} 

\usepackage{bm}
\usepackage{amsmath,stmaryrd,MnSymbol}
\usepackage{graphicx}
\usepackage{hyperref}
\usepackage{setspace}
\usepackage{soul}
\usepackage{accents}
\usepackage{marvosym}
\usepackage{color}
\usepackage{empheq}
\newcommand{\beq}{\begin{equation}}
\newcommand{\eeq}{\end{equation}}
\newcommand{\uu}{{ \bm u}}
\newcommand{\vv}{{ \bm v}}
\newcommand{\rr}{{ \bm r}}
\newcommand{\evec}[1]{\hat{\bm{e}}_{#1}}

\newcommand{\strain}{E}
\newcommand{\stress}{\sigma}
\newcommand{\pressure}{{p}} 
\newcommand{\pone}{{\bf P}^{[1]}} 
\newcommand{\ptwo}{{\bf P}^{[2]}} 
\newcommand{\LegP}{\text{P}}

\begin{document}

\title{Swimmer-Types of Optimum Surface-Driven Active Particles}
\author{
Rafe Md Abu Zayed\aff{1},
Arezoo M. Ardekani\aff{2},
Amir Nourhani\aff{1,3,4}
\corresp{\email{amir.nourhani@gmail.com}}
}

\affiliation{
\aff{1} Department of Mechanical Engineering, University of Akron, Akron, OH 44325
\aff{2} School of Mechanical Engineering, Purdue University, West Lafayette, IN 47907
\aff{3} Biomimicry Research and Innovation Center (BRIC), University of Akron, Akron, OH 44325
\aff{4} Departments of Biology, University of Akron, Akron, OH 44325
}

\maketitle

\begin{abstract}

An optimal microswimmer with a given geometry has a surface velocity profile that minimizes energy dissipation for a given swimming speed. An axisymmetric swimmer can be puller-, pusher-, or neutral-type depending on the sign of the stresslet strength. We numerically investigate the type of optimal surface-driven active microswimmers using a minimum dissipation theorem for optimum microswimmers. We examine the hydrodynamic resistance and stresslet strength with nonlinear dependence on various deformation modes. The results show that microswimmers with fore-and-aft symmetry exhibit neutral-type behavior, while asymmetrical geometries exhibit pusher-, puller-type or neutral-type behavior, depending on the dominant deformation mode and the nonlinear dependence of the stresslet for an optimum microswimmer on deformation mode and amplitude.
\end{abstract}

\section{Introduction}
Microswimmers are small-scale biological or artificial self-propelled entities that navigate through their environments in a regime dominated by viscous forces, making their motion governed by low Reynolds hydrodynamics. Reciprocal motion cannot lead to net propulsion at low Reynolds numbers, due to the reversibility of governing equations as highlighted in Purcell's scallop theorem~\citep{Purcell}. As a result, the movement of biological microswimmers at this scale, such as bacteria and spermatozoa, often relies on non-reciprocal dynamics of flagella or cilia~\citep{lauga2016bacterial, lauga2009hydrodynamics}. Artificial microswimmers,  inspired by these biological systems, similarly exploit non-reciprocal dynamics via surface activity~\citep{Paxton:2004p183,Cordova2013SoftMat6382}, such as in self-phoretic swimmers~\citep{nourhani2024phoresis,laboraving2016,Yariv-11,Golestanian+07}.

The self-propulsion of a microswimmer is force-free and torque-free. Thus, the leading-order flow field generated by axisymmetric particle motion has the symmetry of a force dipole and is characterized by a second-rank symmetric stresslet tensor~\citep{batchelor1970stress}. The sign of the stresslet strength determines the swimmer type: positive values correspond to pullers (similar to {\it C. reinhardtii} and the biflagellated algae {\it Chlamydomonas}), where the thrust is at the front and the resistive drag by the fluid is at the back. In contrast, negative values of the stresslet strength represent pushers with rear thrust (similar to {\it E. coli} and {\it B. subtilis}) ~\citep{spagnolie2015complex, saintillan2015theory, Ardekani_2012}. A zero value corresponds to neutral swimmers, like Volvox, neither push nor pull fluid in either direction~\citep{Drescher, Pedley}.  Understanding these classifications is crucial for predicting how different swimmers interact with their environments, including interactions with other swimmers, boundaries, or obstacles~\citep{spagnolie2015complex,Doostmohammadi}.

The energetic efficiency of microswimmers is a topic of ongoing research. The propulsion of microswimmers arises from the interplay between thrust generation and the hydrodynamics of the particle's interaction with its viscous environment, which results in energy dissipation. For spherical swimmers, it is well-established that neutral swimmers exhibit minimal viscous dissipation and are considered energetically optimal~\citep{Michelin}. However, deviations from perfect spherical symmetry complicate this understanding, as more complex shapes can alter the balance between thrust generation and energy dissipation. Recent studies, including those using boundary element methods, have demonstrated that non-spherical swimmers may exhibit more efficient propulsion as pushers or pullers~\citep{Guo}. A minimum dissipation theorem for surface-driven active microswimmers sets a lower bound on the energy dissipation rate for rigid, surface-driven microswimmers in the regime of low Reynolds number hydrodynamics~\citep{nasouri2021minimum}. Using this theorem, Daddi-Moussa-Ider {\it et al.}~\citep{daddi-moussa-ider_nasouri_vilfan_golestanian_2021} demonstrated that for axisymmetric slightly deformed spheres, only the sign of the third Legendre mode of the geometry determines the type of the optimum axisymmetric swimmer for a given geometry.

Our study builds on these foundations by numerically exploring how the shape of a microswimmer influences its optimal propulsion type using the spectral formalism for Stokes flow~\citep{nabil2022spectral}. We show that deviations from sphericity beyond slight deformations involve additional Legendre modes of the geometry in determining the optimum swimmer type. The sign of the third Legendre mode of shape deformation plays a leading role in determining whether a swimmer type is a pusher, puller, or neutral in most of the geometrical parameter space. However, the phase diagram reveals a domain where, for a given sign of the third mode contribution, the optimal swimmer's behavior can vary. In this domain, the swimmer can be a puller, pusher, or neutral depending on the extent of the deformation. Moreover, our analysis shows that the optimum swimmer type of axisymmetric particles with fore-and-aft symmetry is neutral. The proposed approach provides a theoretical framework for designing artificial microswimmers, allowing for the prediction and optimization of swimmer behavior in various fluid environments.

\section{Optimal microswimmers}
We study the motion of an axisymmetric particle with unit speed along the symmetry axis $\hat{\bf e}$ in a fluid of viscosity $\mu$ under the prescribed surface boundary condition `BC.' The spherical coordinate system $(r, \theta, \phi)$ is used to describe the system. Under the no-slip boundary condition, `NS', the velocity field over the particle surface $\rr_{\!S}$ is 
\beq
\bm{v}_\text{NS}(\rr_{\!S}) = \hat{\bf e}.
\label{eq:no-slip}
\eeq
The flow field due to the perfect-slip boundary condition, `PS', satisfies
\begin{subequations}
  \begin{empheq}[left={\empheqlbrace}]{alignat=2}
\text{Impermeability condition}: \ \ 
\hat{\bf n} \cdot \bm{v}_\text{PS}(\rr_{\!S}) = \hat{\bf n} \cdot \hat{\bf e},\quad\qquad\qquad\qquad\quad
    \label{eq:PS1}
     \\
\text{Zero tangential stress}: \qquad 
\ \
\hat{\bf n} \cdot \bm{\sigma}_\text{PS}(\rr_{\!S})  \cdot \hat{\bf t}
\equiv 2 \mu \, \hat{\bf n} \cdot \bm{E}_\text{PS}(\rr_{\!S})  \cdot \hat{\bf t}
=0,
    \label{eq:PS2}
  \end{empheq}
 \label{eq:PS-BC} 
\end{subequations}
where the stress tensor
$
\bm{\stress}
=
-\pressure \bm{I} + 2\mu \bm{\strain}
$
and the strain tensor
$
 \bm{\strain} = \frac{1}{2} \left[(\bm{\nabla}\uu) + (\bm{\nabla}\uu)^T\right]
$ 
are evaluated over the particle surface, $\hat{\bf n}$ and $\hat{\bf t} = \hat{\bf e}_{\phi} \times \hat{\bf n}$ are the normal and tangent unit vectors on the axisymmetric particle surface, and $p$ is the pressure field.

The minimum dissipation theorem~\citep{nasouri2021minimum} shows that the velocity field $\bm{v}_\text{OM}$ around an \ul{o}ptimum \ul{m}icroswimmer, `OM', moving with unit velocity along its symmetry axis, can be expressed as a linear combination of the velocity fields due to the motion of a passive particle of the same geometry moving with the same velocity under no-slip $\bm{v}_\text{NS}$ and perfect-slip $\bm{v}_\text{PS}$ boundary conditions
\beq
\bm{v}_{\text{OM}}(\rr)
= 
\left(\frac{R_{\text{NS}}}{R_{\text{NS}} - R_{\text{PS}}} \right)
\bm{v}_{\text{PS}}(\rr)
- 
\left(\frac{R_{\text{PS}}}{R_{\text{NS}} - R_{\text{PS}}} \right)
\bm{v}_{\text{NS}}(\rr),
\label{eq:optimumFlowMain1}
\eeq
where the resistance coefficient $R_\text{BC}$ is numerically equal to the force $F_\text{BC} = \hat{\bf e} \cdot {\bm F}_\text{BC}$ exerted by the particle with boundary condition BC = NS, PS on the fluid along $ \hat{\bf e}$.

The multipole expansion of the flow field around a particle $\bm{v}_\text{BC} = \bm{v}_\text{BC}^\text{pf} + \bm{v}_\text{BC}^\text{fd} + \mathcal{O}(r^{-3})$ can be written in terms of increasing powers of $r^{-1}$, where the first two terms exhibit the symmetries of a \ul{p}oint \ul{f}orce and a \ul{f}orce-\ul{d}ipole, respectively. The axisymmetric flow with force-dipole symmetry is characterized by a trace-free, second-rank stresslet tensor
\beq
\bm{S}_\text{BC} = S_\text{BC} \left(\hat{\bf e}\hat{\bf e} - \frac{1}{3} \bm{I}\right),
\eeq
where $S_\text{BC}$ is the strength of the stresslet, and the resulting flow field $\bm{v}_\text{BC}$ is
\beq
\vv_{\text{BC}}(r,\theta)
= 
\frac{-3}{8\pi \mu} \frac{1}{r^2} (\hat{\bf e}_r 
\cdot \bm{S}_\text{BC}  \cdot 
\hat{\bf e}_r) \hat{\bf e}_r
=
\left(\frac{-S_\text{BC}}{4\pi \mu}\right) \frac{1}{r^2} \LegP_2(\cos\theta)\, \hat{\bf e}_r.
\label{eq:optimumFlowMain2}
\eeq
Here, $\hat{\bf e}_r = \rr/r$ and $\LegP_\ell$ is the Legendre polynomial of degree $\ell$. Using the equations for the velocity field around an optimum active particle~(\ref{eq:optimumFlowMain1}), and the linearity of Stokes flow, we have the stresslet strength for the optimum microswimmer moving with unit velocity $\hat{\bf e}$ along its symmetry axis, 
\begin{align}
S_\text{OM}
&= 
\frac{
R_\text{NS}S_\text{PS} 
-
R_\text{PS}S_\text{NS}
}
{R_\text{NS}  - R_\text{PS}}. 
\label{eq:stressletOMMain}
\end{align}

Positive, negative, and zero values of $S_\text{OM}$ correspond to puller, pusher, and neutral swimming type, respectively.
 Daddi-Moussa-Ider {\it et al.} \citep{daddi-moussa-ider_nasouri_vilfan_golestanian_2021} studied the effect of geometry on the swimming mode of optimal swimmers represented by slightly deformed spheres with 
\beq
r_{\!S}(\theta) = r_0 [1 + \delta \xi(\theta)],
\qquad
\xi(\theta) = \sum_{n=1}^\infty \gamma_n \LegP_n(\cos\theta),
\label{eq:deformation}
\eeq
where $\delta$ is the deformation amplitute and the deformation function $\xi$ is expanded in terms of Legendre polynomials. They showed that for slight deformations to the linear order in the deformation amplitude, only mode $n = 3$ of the Legendre polynomials contributes to the stresslet given by Eq.~(\ref{eq:stressletOMMain}),
\beq
S_\text{OM} = \left[8\pi\! \left(\frac{27}{14}\right)\! \mu r_0^2 V_{\!\text{A}}\right] \gamma_3\, \delta + \mathcal{O}(\delta^2). 
\eeq
Thus, optimal swimmers with spherical or slightly deformed spherical geometries, where $\gamma_3 = 0$, are neutral, and the swimming mode of other slightly deformed spheres depends on the sign of $\gamma_3$.

In our study, we numerically evaluate the hydrodynamic resistance and stresslet beyond the linear term in $\delta$ using the spectral method~\citep{nabil2022spectral} and show that optimal microswimmers with fore-and-aft symmetry, corresponding to individual or combinations of even $n$ geometry modes, are neutral. For individual odd values, $n = 3$ has the highest contribution, and the stresslet strength due to $n = 5$ and $n = 7$ is one and two orders of magnitude smaller than that of $n = 3$, respectively. Their swimming modes are opposite to that of $n = 3$. We also evaluate a range of geometries with non-zero $\gamma_3$ and $\gamma_5$ to study the phase diagram of their optimum swimmer types and neutral behavior with non-zero $\gamma_3$.

\section{Calculating Force and Stresslet Using Spectral Method}
The flow field around a particle moving at low Reynolds number in a fluid of viscosity $\mu$ satisfies the Stokes and continuity equations,
\beq
\mu \nabla^2 \bm{u} = \bm{\nabla} \pressure,
\qquad
\bm{\nabla}  \cdot \bm{u} = 0,
\eeq
respectively.
The pressure field satisfies $\nabla^2 \pressure = 0$ and is either a harmonic function or constant. Imposing homogeneity in the radial coordinate, we have Mode-1 biharmonic velocity fields ($\ell \geq 1$) and Mode-2 harmonic velocity fields ($\ell \geq 0$),
\begin{align}
& \text{Mode-1}: \ \nabla^2 \nabla^2 \uu_\ell^{[1]} = 0
\quad \uu_\ell^{[1]}(\rr) = \left(\frac{r_0}{r}\right)^{\ell} \! \uu_\ell^{[1]}(\rr_0),
\quad \quad \ 
\pressure_\ell^{[1]} \propto r^{-(\ell+1)} \LegP_\ell(\cos\theta)
\nonumber \\
& \text{Mode-2}: \ \nabla^2 \uu_\ell^{[2]} = 0 \quad
\quad \uu_\ell^{[2]}(\rr)  = \left(\frac{r_0}{r}\right)^{\ell+2} \! \uu_\ell^{[2]}(\rr_0),
\ \ \quad \pressure_\ell^{[2]} = \text{constant}
\label{eq:StokesModes}
\end{align} 
where $r_0$ is the radius of an arbitrary reference sphere $\mathbb{S}_0$, co-centered with the particle, and  $\uu_\ell^{[\beta]}(\rr_0)$'s are $\theta$-dependent vectorial basis functions defined over $\mathbb{S}_0$. The explicit forms of these modes over the reference sphere are~\citep{nabil2022spectral}:
$\uu_\ell^{[1]}(\rr_0) = [ \ell(\ell+1)\pone_\ell - (\ell-2)\ptwo_\ell]$ and $\uu_\ell^{[2]}(\rr_0) = [-(\ell+1)\pone_\ell + \ptwo_\ell]$ where $\pone_\ell = \LegP_\ell(\cos\theta) \evec{r}$ and $\pone_\ell = \partial_\theta \LegP_\ell(\cos\theta) \evec{\theta}$ are vectorial basis functions.

We can expand the velocity field $\uu(\rr_0)$ over $\mathbb{S}_0$ in terms of $\uu_\ell^{[\beta]}(\rr_0)$. To project onto individual modes, for two vectorial functions $\bm{H}$ and $\bm{G}$ over $\mathbb{S}_0$, we define  
\beq
\langle \bm{H} | \bm{G} \rangle = \int_0^\pi  \bm{H}(\theta) \cdot \bm{G}(\theta)\, \sin\theta \, d\theta = (4 \pi r_0^2)^{-1} \int_{\mathbb{S}_0} \bm{H}(\theta) \cdot \bm{G}(\theta)\, dS
\label{eq:InnerProductVector}
\eeq
 as the inner product.
Moreover, since the Stokes modes~(\ref{eq:StokesModes}) are not orthogonal over the reference sphere,
$\langle   {\bm u}_{\ell}^{[1]}(\rr_0) |  \uu_{\ell}^{[2]} (\rr_0)\rangle \neq 0$, we define their corresponding dual vectors
\begin{align}
&{\bm D}_\ell^{[1]}(\cos\theta) = \frac{2\ell+1}{4 \ell(\ell+1)}\left[ \ell\pone_\ell(\cos\theta) + \ptwo_\ell(\cos\theta)\right]
\\
&{\bm D}_\ell^{[2]}(\cos\theta) = \frac{2\ell+1}{4(\ell+1)}\left[ (\ell-2)\pone_\ell(\cos\theta) + \ptwo_\ell(\cos\theta)\right]
\end{align}
  that satisfy  
$
\left\langle   {\bm D}_{\ell_1}^{[\beta_1]} \middle|  \uu_{\ell_2}^{[\beta_2]} (\rr_0)\right\rangle
=\delta_{\ell_1\ell_2}\delta_{\beta_1\beta_2}
$
where the Kronecker delta $\delta_{ij}$ equals 1 if $i = j$ and 0 otherwise. 

Knowing the velocity field $\uu(\rr_0)$ over the surface of the reference sphere ${\mathbb{S}_0}$, we can write the velocity field in its spectral expansion
\beq
\uu(\rr) 
=
\sum_{\alpha,\ell}
\left\langle \bm{D}_\ell^{[\beta]} \middle| \uu(\rr_0) \right\rangle
\uu_\ell^{[\beta]} (\rr).
\label{eq:SpectralExpansionVelocity}
\eeq
Due to the linearity of Stokes flow and the properties of a Newtonian fluid, a similar spectral expansion with the same expansion coefficients can be written for pressure, strain tensor, and stress tensor.
An axisymmetric particle moving with unit velocity $\hat{\bf e}$ under boundary condition BC exerts a force on the fluid, given by,
\beq
F_\text{BC} =  8\, \pi  \mu r_0 \left\langle \bm{D}_1^{[1]} \middle| \bm{v}_\text{BC}(\rr_0) \right\rangle \stackrel{N}{=} R_\text{BC}
\label{eq:forceBC}
\eeq
where `$N$' atop the equals sign serves as a reminder that the resistance coefficient $R_\text{BC}$ is numerically equal to the magnitude of the force exerted on the fluid by a particle moving with unit velocity. We will use expression (\ref{eq:forceBC}) in Eq.~(\ref{eq:stressletOMMain}) for no-slip and perfect-slip boundary conditions.
Since the velocity field due to the stresslet, according to Eq.~(\ref{eq:optimumFlowMain2}), scales as $r^{-2}$, and the particle is rigid with no fluid source, for $\ell \geq 1$, the only Stokes mode that gives a $\sim r^{-2}$ dependence is Mode-1 with $\ell = 2$. Therefore, the velocity field contribution with the symmetry of a force dipole for a particle moving with unit velocity $\hat{\bf e}$ under boundary condition BC is
\beq
\left\langle \bm{D}_2^{[1]} \middle| 
 \bm{v}_\text{BC}(\rr_0)
\right\rangle
\uu_2^{[1]}(\rr) 
= 
6 r_0^2 \left\langle \bm{D}_2^{[1]} \middle| 
 \bm{v}_\text{BC}(\rr_0)
\right\rangle
\,\frac{1}{r^2} \LegP_2(\cos\theta)\, \evec{r}.
\eeq
Comparing this expression with the flow field~(\ref{eq:optimumFlowMain2}) gives the stresslet strength,
\beq
S_\text{BC} 
= 
-24\, \pi  \mu r_0^2 \left\langle \bm{D}_2^{[1]} \middle| 
 \bm{v}_\text{BC}(\rr_0) 
\right\rangle.
\label{eq:stressletCoeff}
\eeq

After using the expression for the force exerted on the fluid~(\ref{eq:forceBC}) and the stresslet strength~(\ref{eq:stressletCoeff}),
we can rewrite the stresslet strength~(\ref{eq:stressletOMMain}) for the optimum microswimmer moving with unit velocity $\hat{\bf e}$ along its symmetry axis, 
\begin{align}
S_\text{OM}
&= 
-24\, \pi  \mu r_0^2
\left[
\frac{
 \left\langle \bm{D}_1^{[1]} \middle| \vv_\text{NS}(\rr_0) \right\rangle
 \left\langle \bm{D}_2^{[1]} \middle| \vv_\text{PS}(\rr_0) \right\rangle
 -
   \left\langle \bm{D}_1^{[1]} \middle| \vv_\text{PS}(\rr_0) \right\rangle  
  \left\langle \bm{D}_2^{[1]} \middle| \vv_\text{NS}(\rr_0) \right\rangle  
}{
 \left\langle \bm{D}_1^{[1]} \middle| \vv_\text{NS}(\rr_0) \right\rangle
 -
  \left\langle \bm{D}_1^{[1]} \middle| \vv_\text{PS}(\rr_0) \right\rangle
}
\right]
\label{eq:stressletOM}
\end{align}
which can be calculated using the $\ell=1, 2$ coefficients of Mode-1 terms in the spectral expansion~(\ref{eq:SpectralExpansionVelocity}) for flow fields with no-slip (\ref{eq:no-slip}) and perfect-slip (\ref{eq:PS-BC}) boundary conditions, as elaborated below.

To calculate the unknown expansion coefficients $\langle \bm{D}_{\ell}^{[\beta]} | \bm{v}_\text{NS}(\rr_0) \rangle$ for an axisymmetric particle with surface $\mathbb{S}$, which for a non-spherical particle is different from $\mathbb{S}_0$, we apply the no-slip boundary condition~(\ref{eq:no-slip}) over $\mathbb{S}$ to the spectral expansion~(\ref{eq:SpectralExpansionVelocity}) and take the inner product with the dual vectors, yielding
\beq
\left\langle 
\bm{D}_{\ell_1}^{[\beta_1]}
\middle|
\bm{v}_\text{NS}(\rr_{\!S})
\right\rangle
=
\sum_{\beta_2,\ell_2}
\left\langle 
\bm{D}_{\ell_1}^{[\beta_1]}
\middle|
\uu_{\ell_2}^{[\beta_2]} (\rr_{\!S})
\right\rangle
\left\langle \bm{D}_{\ell_2}^{[\beta_2]} \middle| 
\bm{v}_\text{NS}(\rr_0)
\right\rangle.
\label{eq:noslip-inline}
\eeq 
Here, except for the expansion coefficients, the rest of the terms are known.

For the perfect-slip scenario, both sides of Eqns.~(\ref{eq:PS-BC}) are scalar functions of $\theta$. We define the inner product  
$\langle h | g \rangle = \int_0^\pi  h(\theta) \, g(\theta)\, \sin\theta \, d\theta$
for two scalar functions, 
which is distinct from the inner product of two vectorial functions (\ref{eq:InnerProductVector}), although the same notion applies. Thus, using inner products with Legendre polynomials, we can turn the perfect-slip boundary condition~(\ref{eq:PS-BC}) into a set of linear equations with expansion coefficients as unknowns
\begin{subequations}
\begin{align}
\left\langle 
\LegP_{\ell_1}
\middle| 
\hat{\bf n}  \cdot \bm{v}_\text{PS}(\rr_{\!S})
\right\rangle
&=
\sum_{\beta_2,\ell_2}
\left\langle 
\LegP_{\ell_1}
\middle| 
\hat{\bf n}  \cdot \uu_{\ell_2}^{[\beta_2]} (\rr_{\!S})
\right\rangle
\left\langle \bm{D}_{\ell_2}^{[\beta_2]} \middle|
\bm{v}_\text{PS}(\rr_0) 
 \right\rangle
\\
0
&=
\sum_{\beta_2,\ell_2}
\left\langle 
\LegP_{\ell_1}
\middle| 
\hat{\bf n}  \cdot \bm{\strain}_{\ell_2}^{[\beta_2]}(\rr_{\!S}) \cdot \hat{\bf t} 
\right\rangle
\left\langle \bm{D}_{\ell_2}^{[\beta_2]} \middle| 
\bm{v}_\text{PS}(\rr_0)
\right\rangle,
\label{eq:perfectslip-inline}
\end{align}
\label{eq:perfectslip-inline}
\end{subequations}
where the strain tensor for Stokes modes is
$
 \bm{\strain}_\ell^{[\beta]} = \frac{1}{2} [(\bm{\nabla}\uu_\ell^{[\beta]}) + (\bm{\nabla}\uu_\ell^{[\beta]})^T]
$.

In the next section, by obtaining the expansion coefficients $\langle \bm{D}_\ell^{[\beta]} | \vv_\text{BC}(\rr_0) \rangle$ from Eqns.~(\ref{eq:noslip-inline}) and (\ref{eq:perfectslip-inline}) for no-slip and perfect-slip boundary conditions, respectively, we will calculate the stresslet strength~(\ref{eq:stressletOM}) for optimum microswimmers for a set of geometries and discuss their swimming type.

\begin{figure}
    \centering
    \includegraphics[width = \textwidth]{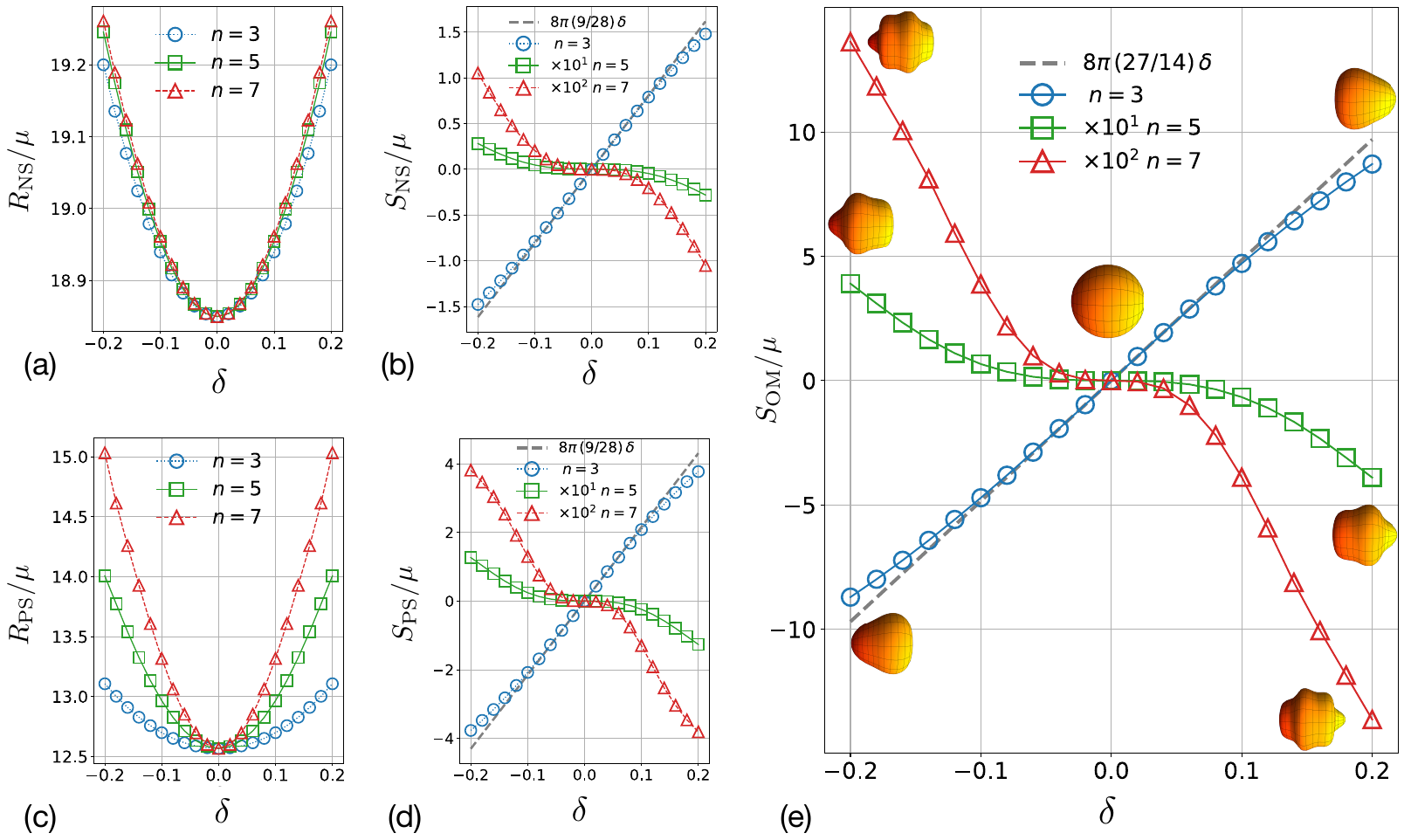}
    \caption{
    The force and stresslet strength for particles with individual geometry modes $n = 3, 5, 7$ under (a)\&(b) no-slip boundary condition and (c)\&(d) perfect-slip boundary conditions. (e) The stresslet for the optimum swimmer. In these calculations $r_0 =1$.}
    \label{fig:fig1}
\end{figure}

\section{Results and Discussion}

For a particle with fore-and-aft symmetry, $\bm{D}_2^{[1]} \cdot \bm{v}_\text{BC}(\rr_0)$ is odd with respect to $\cos\theta$, and thus, the stresslet strength is zero, $S_\text{BC} \propto \left\langle \bm{D}_2^{[1]} \middle| \bm{v}_\text{BC}(\rr_0) \right\rangle = 0$. Consequently, the stresslet strength for any optimum swimmer with a geometry consisting of individual even geometrical modes $\gamma_{2k}$ with $k \in \mathbb{N}$, or linear combinations of these even modes, is zero. Therefore, the optimum swimmer type for a particle with fore-and-aft symmetry is neutral.

For individual odd geometrical modes $n = 3, 5$, and 7, for which $\xi(\theta) = \LegP_n(\cos \theta)$, we evaluated the force and stresslet strength for particles under no-slip (Fig. \ref{fig:fig1}(a)\&(b)) and perfect-slip (Fig. \ref{fig:fig1}(c)\&(d)) boundary conditions, and calculated the stresslet for the optimum swimmer (Fig. \ref{fig:fig1}(e)) using Eq.~(\ref{eq:stressletOM}) for a range of deformation amplitudes $\delta \in [-0.2, 0.2]$. For $n = 3$ ($\gamma_n =\delta_{n3}$ in Eq.~(\ref{eq:deformation})), the numerical values of the stresslet strength corroborate the linear theoretical values: $S_\text{NS} = 8 \pi \left(\frac{9}{28}\right) \mu r_0^2 \,\delta$ for no-slip, $S_\text{PS} = 8 \pi \left(\frac{6}{7}\right) \mu r_0^2 \,\delta$ for perfect-slip, and $S_\text{OM} = 8 \pi \left(\frac{27}{14}\right) \mu r_0^2 \,\delta$ for optimum swimmers~\citep{daddi-moussa-ider_nasouri_vilfan_golestanian_2021} in the range $|\delta| \lesssim 0.1$, with slight deviations in the range $0.1 \lesssim  |\delta| \leq 0.2$.

For particles with $n = 3$ geometry mode, positive values of the deformation amplitude $\delta$ correspond to pullers, while the optimum swimmer type for negative values of $\delta$ corresponds to pushers. For $n = 5$ and $7$, the sign of the stresslet is opposite to that of $n = 3$. Therefore, the swimmer-type behavior is reversed:
 the optimum swimmer type for positive (negative) values of $\delta$ corresponds to pushers (pullers). Moreover, the stresslet for $n = 5$ is one order of magnitude smaller and for $n = 7$ two orders of magnitude smaller than that of $n = 3$, making $n = 3$ the dominant geometrical mode. As shown in Fig.~\ref{fig:fig1}(e), the slope of the stresslet strength for $n = 5$ and $7$ is effectively zero at $\delta = 0$. Thus, in the linear regime of very small $\delta$, these modes do not contribute to determining the swimmer type and only the contribution from $n = 3$ appears, while the rest of the geometry modes ($n \neq 3$) are neutral. However, as the figure shows, in the nonlinear regime, other odd $n$ geometry modes are not neutral.

\begin{figure}
    \centering
    \includegraphics[width = \textwidth]{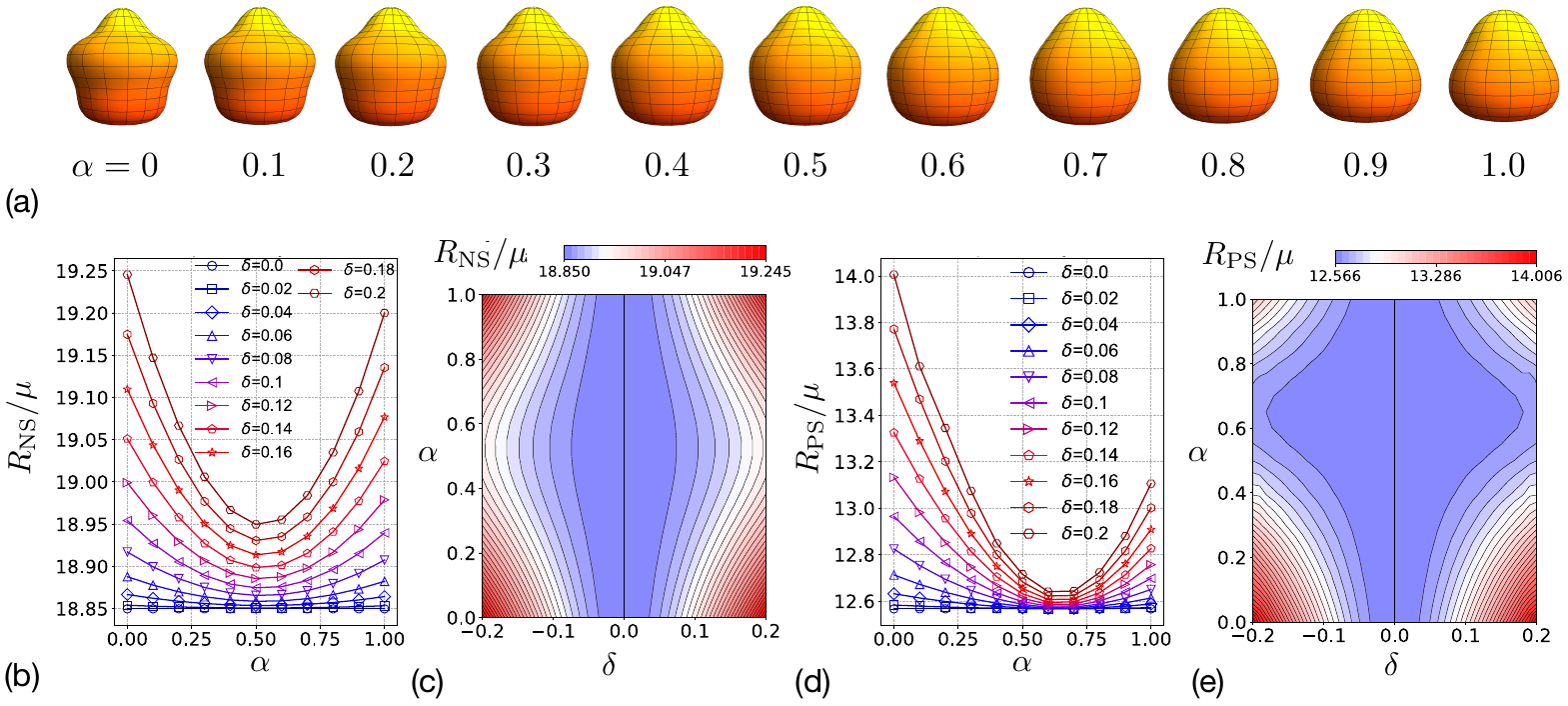}
    \caption{
(a) Particle geometries corresponding to the deformation function $\xi_{35}(\theta; \alpha) = \alpha \LegP_3 + (1-\alpha) \LegP_5$ for $\alpha \in [0,1]$ and $\delta =0.2$.
Resistance coefficients for (b)\&(c)
no-slip and (d)\&(e) perfect-slip boundary conditions as functions of $\alpha$ and $\delta$.
    }
    \label{fig:fig2}
\end{figure}

\begin{figure}
    \centering
    \includegraphics[width = \textwidth]{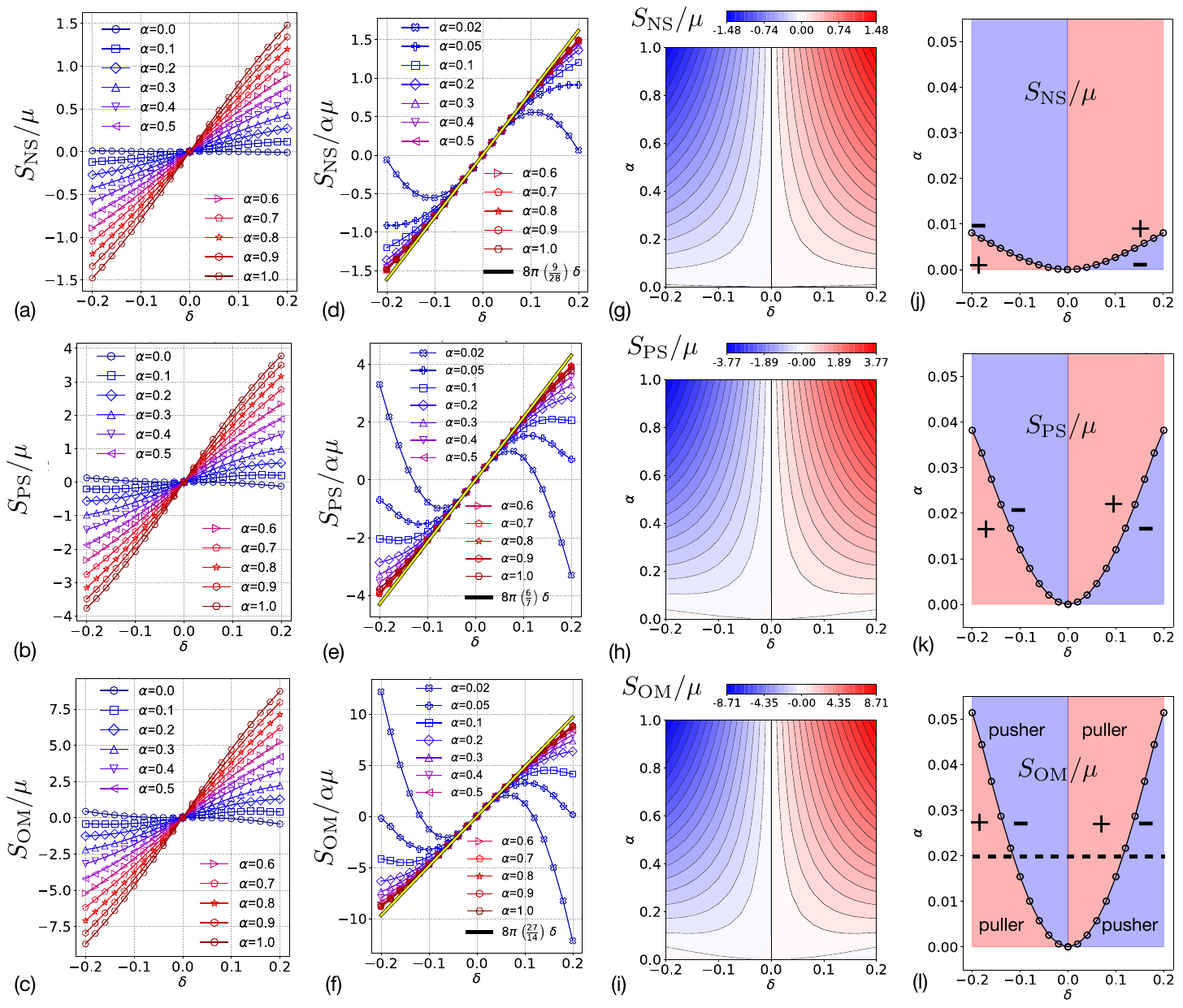}
    \caption{(a)-(c) Stresslet strength corresponding to the deformation function $\xi_{35}(\theta; \alpha) = \alpha \LegP_3 + (1-\alpha) \LegP_5$ for $\alpha \in [0,1]$ and $\delta \in [-0.2, 0.2]$ for no-slip, perfect-slip, and optimum swimmer boundary conditions, respectively. (d)-(f) Scaled stresslet strength based on the contribution of the third Legendre mode $\alpha \equiv \gamma_3$. (g)-(i) Contour plots of the stresslet strength. (j)-(l) Phase diagram showing the sign of the stresslet strength.}
    \label{fig:fig3}
\end{figure}

Next, we examine the deformation function $\xi_{35}(\theta; \alpha)$ defined by
\beq
\xi_{nm}(\theta; \alpha) = \alpha \LegP_n(\cos \theta) + (1 - \alpha) \LegP_m(\cos \theta), 
\qquad 
\alpha \in [0,1],
\eeq
such that $\xi_{35}(\theta; 0) = \LegP_5(\cos\theta)$ and $\xi_{35}(\theta; 1) = \LegP_3(\cos\theta)$ as shown in Fig.~\ref{fig:fig2}(a) for $\delta = 0.2$. This particular shape function ($\gamma_3 = \alpha$ and $\gamma_5 = 1-\alpha$) allows us to explore the interplay between the geometrical modes $n = 3$ and $n = 5$ in determining the optimum swimmer type of the active particle. In the domain of $\delta >0$, $n = 3$ acts as a puller and $n = 5$ as a pusher, and the behavior is reversed for opposite signs of $\delta<0$, making the function $\xi_{35}(\theta; \alpha)$ useful for studying transitions between pusher and puller types; varying the parameter $\alpha$ enables a continuous shift between swimmer types, depending on which geometry mode dominates for a given value of $\alpha$ and $\delta$.

As shown in Fig.~\ref{fig:fig2}, the variation in the resistance coefficient with respect to $\alpha$ is not linear, indicating a nonlinear interaction between the geometry modes that goes beyond a simple linear weighted average. This nonlinearity suggests that the transition between pusher- and puller-type behavior involves intricate changes in the fluid dynamics around the particle, influenced by both the relative contributions of the geometrical modes and the deformation amplitude $\delta$. In Fig.~\ref{fig:fig3}(a), (b), and (c), we present the calculations for the stresslet across various values of $\alpha$ in the range \mbox{$\delta \in [-0.2, 0.2]$}. Except for the case of $\alpha = 0$, which corresponds to the $n = 5$ mode, the stresslet exhibits a monotonic increase with increasing $\delta$ for the rest of the $\alpha$ values presented in the figure.

Since $\alpha \equiv \gamma_3$ in Eq.~(\ref{eq:deformation}), to evaluate the contribution of the geometrical mode $n=3$, we plotted $S_\text{BC}/\alpha$ for $\alpha \neq 0$ to examine the relationship between the stresslet and deformation amplitude $\delta$ in Fig.~\ref{fig:fig3}(d), (e), and (f). As expected, we observe a linear relationship with $\delta$ for small values of $|\delta| \lesssim 0.05$, and beyond this domain, deviations from linearity begin to emerge, indicating the influence of $n = 5$ on the stresslet. By analyzing $S_\text{OM}/\alpha$, we can further understand the impact of this secondary geometrical mode on the particle's swimming dynamics. 

Furthermore, we created a phase diagram in the regime of small $\alpha$ values to explore the transition between different swimmer types, as shown in Fig.~\ref{fig:fig3}(j), (k), and (l). This phase diagram reveals the boundary where, for a given deformation amplitude $\delta$, the swimmer becomes neutral—meaning the stresslet strength is zero for non-zero value of $\gamma_3$. Above this boundary curve, the swimmer predominantly exhibits the swimmer type corresponding to the majority contribution from the $n = 3$ mode. Below the curve, in a narrow range of $\alpha$ values, the swimmer displays the opposite type behavior, corresponding to the dominance of the $n = 5$ mode. Therefore, for a given sign of $\gamma_3$ the swimmer can be puller, pusher, or neutral depending on the extent of deformation $\delta$. The phase diagram provides a detailed view of how the competition between geometrical modes affects the overall swimmer type behavior.

\section{Conclusion}
Our study presents a numerical analysis of the swimmer types of optimum surface-driven active particles, focusing on the interplay between the geometrical modes. Utilizing the minimum dissipation theorem, we systematically evaluated the contributions of various geometrical modes to the swimming type of deformed spherical microswimmers. Our findings indicate that microswimmers with fore-and-aft symmetry exhibit neutral swimming type due to the cancellation of stresslet contributions. In contrast, asymmetrical geometries display distinct pusher- or puller-type behaviors depending on the dominant deformation mode and the nonlinear dependence of the stresslet on the deformation amplitude. The phase diagram constructed for small third Legendre mode of deformation reveals the sensitivity of the optimum swimmer type to geometric perturbations, providing insight into the transition between pusher and puller behavior. These results contribute to a deeper understanding of the intricate relationship between geometry and swimming efficiency in active microswimmers, which could inform the design of artificial microswimmers for practical applications.\\

\noindent
{\bf Acknowledgements.}
We extend our sincere gratitude to Paul E. Lammert for his insightful discussions and valuable comments. \\

\noindent
{\bf Funding.}
RZ and AN acknowledge support from the National Science Foundation CAREER award, grant number CBET-2238915. AMA acknowledges support from the National Science Foundation, grant number CBET-2341154.\\

\noindent
{\bf Declaration of interests.} The author reports no conflict of interest.


\end{document}